\documentclass[10pt,letterpaper]{article}
\usepackage{opex3}
\usepackage{cite}
\usepackage{amsmath,amssymb,mathrsfs}
\usepackage{indentfirst}
\begin{document}

\title{Robustness via Diffractal Architectures}

\author{Matthew Moocarme$^{1,2}$ and Luat T. Vuong$^{1,2,*}$}

\address{$^1$The Graduate Center of The City University of New York, 365 5th Ave New York, NY, 10016\\
$^2$Queens College of The City University of New York, 65-30 Kissena Blvd, Flushing, NY, 11367}

\email{$^*$Luat.Vuong@qc.cuny.edu} 



\newsavebox{\smlmat}
\savebox{\smlmat}{$\left(\begin{smallmatrix}2&-2\\-1&2\end{smallmatrix}\right)$}

\begin{abstract}
When plane waves diffract through fractal-patterned apertures, the resulting far-field profiles or diffractals also exhibit iterated, self-similar features. Here we show that this specific architecture enables robust signal processing and spatial multiplexing: arbitrary parts of a diffractal contain sufficient information to recreate the entire original sparse signal.
\end{abstract}


\bibliographystyle{unsrt}
\bibliography{DiffractalBib_30Jul}

\begin{thebibliography}{10}

\bibitem{Mandlebrot}
B.~B. Mandlebrot.
\newblock {\em The Fractal Geometry of Nature}.
\newblock Freeman, New York, 1982.

\bibitem{Sorensen}
C.~M. Sorensen.
\newblock {Light scattering by fractal aggregates: A review}.
\newblock {\em {Aerosol Science and Technology}}, {35}({2}):{648--687}, {Aug}
  {2001}.

\bibitem{Macke}
A.~Macke, J.~Mueller, and E.~Raschke.
\newblock {Single scattering properties of atmospheric ice crystals}.
\newblock {\em {Journal of the Atmospheric Sciences}}, {53}({19}):{2813--2825},
  {Oct} {1996}.

\bibitem{Schmitt}
J.~M. Schmitt and G.~Kumar.
\newblock {Optical scattering properties of soft tissue: a discrete particle
  model}.
\newblock {\em {Applied Optics}}, {37}({13}):{2788--2797}, {May} {1998}.

\bibitem{Soljacic}
M.~Soljacic, M.~Segev, and C.~R. Menyuk.
\newblock {Self-similarity and fractals in soliton-supporting systems}.
\newblock {\em {Phys Rev E}}, {61}({2}):{R1048--R1051}, {Feb} {2000}.

\bibitem{Segev}
M.~Segev, M.~Soljacic, and J.~M. Dudley.
\newblock {Fractal optics and beyond}.
\newblock {\em {Nature Photonics}}, {6}({4}):{209--210}, {Apr} {2012}.

\bibitem{Stockman}
M.~I. Stockman, V.~M. Shalaev, M.~Moskovits, R.~Botet, and T.~F. George.
\newblock Enhanced raman scattering by fractal clusters: Scale-invariant
  theory.
\newblock {\em Phys. Rev. B}, 46:2821--2830, Aug 1992.

\bibitem{Tsai}
D.~P. Tsai, J.~Kovacs, Z.~Wang, M.~Moskovits, V.~M. Shalaev, J.~S. Suh, and
  R.~Botet.
\newblock Photon scanning tunneling microscopy images of optical excitations of
  fractal metal colloid clusters.
\newblock {\em Phys. Rev. Lett.}, 72:4149--4152, Jun 1994.

\bibitem{Berry}
M.~V. Berry.
\newblock {Diffractals}.
\newblock {\em {J. Phys A - Mathematical and General}}, {12}({6}):{781--797},
  {1979}.

\bibitem{Horvath}
P.~Horvath, P.~Smid, I.~Vaskova, and M.~Hrabovsky.
\newblock {Koch fractals in physical optics and their Fraunhofer diffraction
  patterns}.
\newblock {\em {Optik}}, {121}({2}):{206--213}, {2010}.

\bibitem{Hou}
B.~Hou, G.~Xu, W.~J. Wen, and G.~K.~L. Wong.
\newblock {Diffraction by an optical fractal grating}.
\newblock {\em {Applied Physics Letters}}, {85}({25}):{6125--6127}, {Dec}
  {2004}.

\bibitem{Barrera}
John~F. Barrera, Myrian Tebaldi, Dafne Amaya, Walter~D. Furlan, Juan~A.
  Monsoriu, Nestor Bolognini, and Roberto Torroba.
\newblock {Multiplexing of encrypted data using fractal masks}.
\newblock {\em {Optics Letters}}, {37}({14}):{2895--2897}, {Jul 15} {2012}.

\bibitem{Unnikrishnan}
G~Unnikrishnan, J~Joseph, and K~Singh.
\newblock {Optical encryption by double-random phase encoding in the fractional
  Fourier domain}.
\newblock {\em {Optics Letters}}, {25}({12}):{887--889}, {Jun 15} {2000}.

\bibitem{Verma}
R.~Verma, M.~K. Sharma, V.~Banerjee, and P.~Senthilkumaran.
\newblock {Robustness of Cantor Diffractals}.
\newblock {\em {Optics Express}}, {21}({7}):{7951--7956}, {Apr} {2013}.

\bibitem{Verma2}
R.~Verma, V.~Banerjee, and P.~Senthilkumaran.
\newblock {Redundancy in Cantor Diffractals}.
\newblock {\em {Optics Express}}, {20}({8}):{8250--8255}, {Apr} {2012}.

\bibitem{Kelly}
M.~F. Duarte, M.~A. Davenport, M.~B. Wakin, J.~N. Laska, D.~Takhar, K.~F.
  Kelly, and R.~G. Baraniuk.
\newblock Multiscale random projections for compressive classification.
\newblock In {\em Image Processing, 2007. ICIP 2007. IEEE International
  Conference on}, volume~6, pages VI -- 161--VI -- 164, Sept 2007.

\bibitem{Howland}
Gregory~A. Howland and John~C. Howell.
\newblock {Efficient High-Dimensional Entanglement Imaging with a
  Compressive-Sensing Double-Pixel Camera}.
\newblock {\em {Physical Review X}}, {3}({1}), {FEB 20} {2013}.

\bibitem{Radonic}
V.~Radonic, K.~Palmer, G.~Stojanovic, and V.~Crnojevic-Bengin.
\newblock {Flexible Sierpinski Carpet Fractal Antenna on a Hilbert Slot
  Patterned Ground}.
\newblock {\em {International Journal of Antennas and Propagation}}, {2012}.

\bibitem{Puente-Baliarda}
C.~Puente-Baliarda, J.~Romeu, R.~Pous, and A.~Cardama.
\newblock {On the behavior of the Sierpinski multiband fractal antenna}.
\newblock {\em {IEEE Transactions on Antennas and Propagation}},
  {46}({4}):{517--524}, {Apr} {1998}.

\bibitem{Jacquin}
A.~E. Jacquin.
\newblock {Fractal Image-Coding - A Review}.
\newblock {\em {Proceedings of the IEEE}}, {81}({10}):{1451--1465}, {Oct}
  {1993}.

\bibitem{Allouche}
J.~P. Allouce and Shallit.
\newblock {\em Automatic Sequence: Theory, Applications, Generalizations}.
\newblock Cambridge University Press, 2003.

\bibitem{Code1}
M.~Moocarme.
\newblock {Sierpinski carpet generator}.
\newblock
  \url{https://github.com/moocarme/Diffractals/blob/master/Sierpinski.m}, 09
  2015.

\bibitem{Code2}
M.~Moocarme.
\newblock {Sierpinski carpet reconstruction algorithm}.
\newblock \url{https://github.com/moocarme/Diffractals/blob/master/reconFun.m},
  09 2015.

\bibitem{Goodman}
J.~W. Goodman.
\newblock {\em Introduction to Fourier optics}.
\newblock Roberts and Co., 2005.

\end{thebibliography}
\section{Introduction}
Many natural systems exhibit fractal properties~\cite{Mandlebrot}; in fact, scale invariance underlies many self-similar phenomena, from frost crystallization to animal colouration to stock-market pricing. In the realm of optics, the fractal anatomy of systems is widely associated with aggregated dielectric and metal colloids \cite{Sorensen}, crystals \cite{Macke}, and tissues \cite{Schmitt}. Fractal systems are also observed in nonlinear optics~\cite{Soljacic, Segev} and fractalized optical properties can even efficiently characterize or enhance the response of materials~\cite{Stockman, Tsai}. Less well utilized are the features of diffractals, or the diffraction patterns of fractal signals ~\cite{Berry,Horvath,Hou}. Diffractals feature in methods of encrypting data~\cite{Barrera} as a versatile approach to double random-phase encoding~\cite{Unnikrishnan}. However, to differentiate from previous work we exploit fractal architecture to improve transmission robustness.

Here, we explore diffractals for their application in signal processing ~\cite{Verma,Verma2}.  We show that the free-space propagation of diffractal-signal architectures provides algorithmic value and spatial multiplexing properties; any arbitrary subsection of a diffractal contains sufficient information to recreate the original sparse signal that is transmitted with a specific fractal architecture. In a manner similar to compressive imaging \cite{Kelly, Howland}\textemdash where sparse signals reveal greater information via the diffraction through structures\textemdash here the fractal structuring within the signal sparseness prevents the loss of information.

Like other applications of fractals in communications applications, the diffractal architecture exhibits trade-offs. Fractal antennas for the radio frequency and microwave regimes are known for being compact and versatile over wide spectral bands but are power intensive \cite{Radonic,Puente-Baliarda}; fractal encoding algorithms enable image compression with higher-resolution at the expense of greater algorithmic complexity~\cite{Jacquin}; here, our research identifies that diffractal architectures prevent the loss of information but require greater signal preprocessing.  This investigation extends our understanding of fractal structures in signal communications and may increase robustness and transmission rates of satellite, wireless, and interplanetary communication systems, {\it i.e.,} networks that support a large number of roaming receivers.

The remainder of this report is organized as follows.  First, we formalize the form of a transmitted fractal signal, show that its far-field diffraction pattern or diffractal is also a fractal, and illustrate the reciprocal nature of fractals with the Sierpinski carpet. Secondly, we demonstrate the robust retrieval of a signal from a diffractal; the original signal is reconstructed even when the majority of the diffractal signal is blocked. Finally, we discuss the future applications for diffractal spatial multiplexing in free-space communication systems and conclude.

\section{Theoretical Description}\label{theory} 
\subsection{Spatial Multiplexing of the Diffractal} 

We generate the fractal transmittance pattern from any base matrix via recursive iterations where the matrix is resized and convolved with itself repeatedly \cite{Allouche}.  The base matrix $B(x,y)$ adopts a general form,
\begin{equation}
B_i(x,y) = \sum_j\delta(xr^{i-1}-x_i,yr^{i-1}-y_j),
\label{BJ1}
\end{equation}
where the subscript denotes the iteration $i$, $r>1$ is the relative scaling between iterations, and $\delta(x-x_j,y-y_j)$ is the Dirac delta function at $x = x_j$ and $y=y_j$. 
The fractal transmittance function $T(x,y)$ is calculated recursively,
\begin{equation}
T_n(x,y) = T_{n-1}(x,y)*B_{n-1}(x,y),
\label{nTrans}
\end{equation}
where the subscript denotes the order of the fractal or its expression at the $n^{th}$ iteration, $T_0$ is the initial profile of 1's, and $*$ denotes the convolution operator.  

Subsequently, the diffractal is the Fourier transform or the far-field of the transmittance function $\tilde{T}$ ~\cite{Horvath},
\begin{equation}
\tilde{T}_n(k_x,k_y) = \tilde{T}_0(k_x,k_y)\prod_{i=1}^{n}\tilde{B}_i(k_x,k_y)\label{FTn}.
\end{equation}

The Sierpinski carpet is one example of a fractal that is generated by this process and via the iterated substitution of a $3\times 3$ base matrix of ones with removal of the center element:
\begin{equation}
\Bigg\{0\rightarrow \Bigg[\begin{array}{ccc}
0 & 0 & 0 \\
0 & 0 & 0\\
0 & 0 & 0\end{array}\Bigg], 1\rightarrow\Bigg[\begin{array}{ccc}
1 & 1 & 1 \\
1 & 0 & 1\\
1 & 1 & 1\end{array}\Bigg] \Bigg\}.
\label{baseMatrix}
\end{equation}

The second substitution of Eq. \ref{baseMatrix} represents the mathematical expression for the base matrix $B_0(x,y)$.  In the case of the Sierpinski-carpet base-matrix elements, $r=3$, and $x_j$ and $y_j$ are the perimeter coordinates of a $3\times 3$ 9-unit block centered at the origin, and $(x_j,y_j)\in [(1,1),(1,0),(1,-1),(0,-1),(-1,-1),(-1,0),(-1,1),(0,1)]$. Since each of the Dirac delta functions in $B_0(x,y)$ yields a phase shift in the Fourier domain, {\it i.e.,} $\mathscr{F}\left\{\delta(\alpha x-x_i)\right\} = e^{2\pi i_ck_xx_i}/|\alpha|$, its scaled Fourier Transform at the $i^{th}$ iteration of the Sierpinski carpet [Eq. \ref{BJ1}] becomes:
\begin{equation}
\tilde{B}_i(k_x,k_y) = (2/r)^{i-1}[\cos(2\pi3^{1-i}k_x)\cos(2\pi3^{1-i}k_y)+\cos(2\pi 3^{1-i}k_x)+\cos(2\pi 3^{1-i}k_y)].\label{BF}
\end{equation}
With each iteration, the spatial frequency components $k_x,k_y$ of the diffractal increase by a factor of 3 and spread the diffractal across a 3-times wider $k_x,k_y$-range, which is evident in Eq. \ref{BF}; the cutoff of $\tilde{T}_n$ scales in proportion with $k_x, k_y \propto r^{n-1}$. 

\begin{figure}[t!]
 \includegraphics[width=\textwidth]{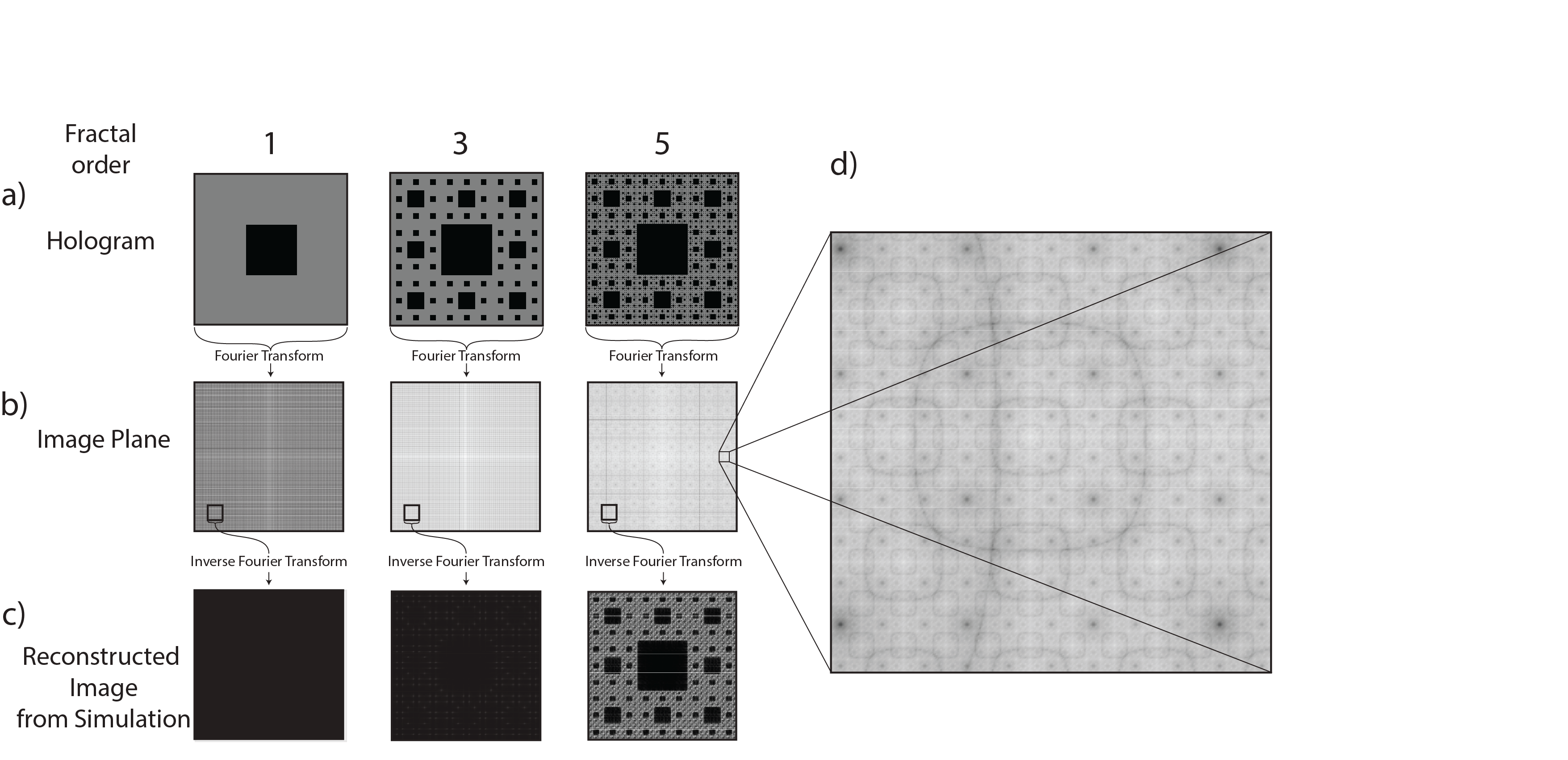}
\caption{a) Signal patterns, or fractalized signals (FS) of orders $n = 1$, $3$, and $5$. b) Corresponding Fourier transforms of signals on logarithm scale, or diffractal signals (DS). c) Reconstructed Fourier-transforms (RFS) from the 1\% black-outlined subset, the blocked diffractal signal (BDS). d) Enlargement of a portion of the $n=5$ diffractal, which illustrates similar patterns at different length scales.}
\label{AnalRecon}
\end{figure}

We calculate the Sierpinski carpet $T$ recursively [Eq. \ref{nTrans}]: a second-order fractal is generated from the Kronecker tensor product of the base matrix $\Big[\begin{smallmatrix} 1 & 1 & 1 \\ 1 & 0 & 1\\ 1 & 1 & 1\end{smallmatrix}\Big]$ with itself; a third-order fractal is generated from the Kronecker tensor product of a second-order fractal and the same base matrix [see Code File 1]~\cite{Code1}. Sierpinski carpets of $n= 1, 3,$ and $5$ are shown in Fig. \ref{AnalRecon}a) with the corresponding diffractals $\tilde{T}$ [Fig. \ref{AnalRecon}b)]. 

As the fractal order increases, $\tilde{T}$ exhibits smaller self-similar features at higher $k_x,k_y$;  the diffractal also exhibits a fractal architecture, as observed in other literature~\cite{Berry,Horvath}.  Moreover, when $n$ is large, an arbitrary subsection of $\tilde{T}$ closely resembles the whole, and the self-similarity is already apparent with $n= 5$ [Fig. \ref{AnalRecon}c-d)]. The iterated, self-similar, wide-spatial-frequency features contained in the diffractal $\tilde{T}_n$ [Eq. \ref{FTn}] enable robust reconstruction of itself, which is described in the next subsection.

\subsection{Robust Reconstruction of a Blocked Diffractal Signal} 

We now refer to the sparse matrix $B_0$ and transmitted signal $T$ as the original and fractalized signal, OS and FS, respectively; the diffractal signal DS is $\tilde{T}$ or the Fourier transform of FS; BDS refers to an off-axis subsection of DS that is filtered; a reconstructed fractalized signal RFS refers to the inverse-Fourier transform of BDS; a regenerated version of the original signal ROS interpolates RFS in order to obtain OS.  

In Fig. \ref{AnalRecon}b), a subsection or BDS is outlined with a black square, which represents 1\% of DS. The corresponding RFS from BDS are shown in Fig. \ref{AnalRecon} c). For $n$ = 1, 3, RFS is ostensibly blank because BDS carries negligible power.  In contrast, when $n=$5, RFS carries features that resemble FS.  We refer to the capacity to reproduce FS with 1\% of the off-axis  and 0.1\% of the transmitted power from the OS as ${\it robust}$ reconstruction.  

Robust reconstruction\textemdash where RFS resembles FS\textemdash is possible even when the size of BDS is significantly reduced.  A corresponding increase in the FS fractal order $n$ will yield RFS that resembles FS when BDS is arbitrarily small. With the example of the Sierpinski carpet [Fig. \ref{AnalRecon}], if the size of BDS is reduced to 0.1\% of OS, then a comparable RFS is produced by increasing the fractal order of FS from $n=5$ to $n = 8$.  

Robust reconstruction also refers to the capacity to regenerate OS from RFS from a simple threshold function [see Code File 2]~\cite{Code2}. The simple threshold function of the Sierpinski carpet divides RFS into a $3\times 3$ array and measures the intensity in each of the 9 elements. Above a certain threshold value, the element is assigned a value of 1, and otherwise assigned a value of 0. The $3\times 3$ array that is processed with the threshold function, ROS, is theoretically identical to OS when OS is transmitted with the diffractal architecture. 

A partial explanation for the robust reconstruction is that increasing-order FS carry arbitrarily-high $k_x,k_y$ and enable arbitrarily-{\it small} BDS to carry the information of FS or OS.  If we strictly limit OS to binary or Dirac-delta functions, then FS has no $k_x,k_y$ cut-off and as $n$ approaches infinity, fractalized features appear in FS without a $k_x,k_y$ cut-off.  In fact, the amplitude of the additional $k_x,k_y$ gained from each iteration scales inversely with $r^{n-1}$, which provides a detection limit only in practice; in theory, each iteration generates high-$k_x,k_y$ copies of OS [Eq. \ref{baseMatrix}] that are spatially distributed from the origin.  

Yet it is worth noting that the robust reconstruction is achieved because the diffractal architecture also couples $k_x$ and $k_y$ in iterated products [Eq. \ref{FTn}].  As a result, RFS will resemble FS when either the BDS size {\it or location} changes.  In a manner similar to spatial filtering, RFS will produce outlines of FS if $n$ is not sufficiently large; however, unlike a high-pass spatial filter of a multi-scale random high-$k_x,k_y$ pattern \cite{Kelly}, a change in the location or the size of BDS will not distort the outline of RFS.  Subsequently, the diffractal architecture provides superior performance over other algorithms that regenerate sparse data after filtering \cite{Kelly, Howland}; any of the reconstructed features in the ROS are strictly limited to the $r \times r$-elements that are nonzero.   


\section{Experimental Results} \label{experiment}
\begin{figure}[t!]
\includegraphics[width=\textwidth]{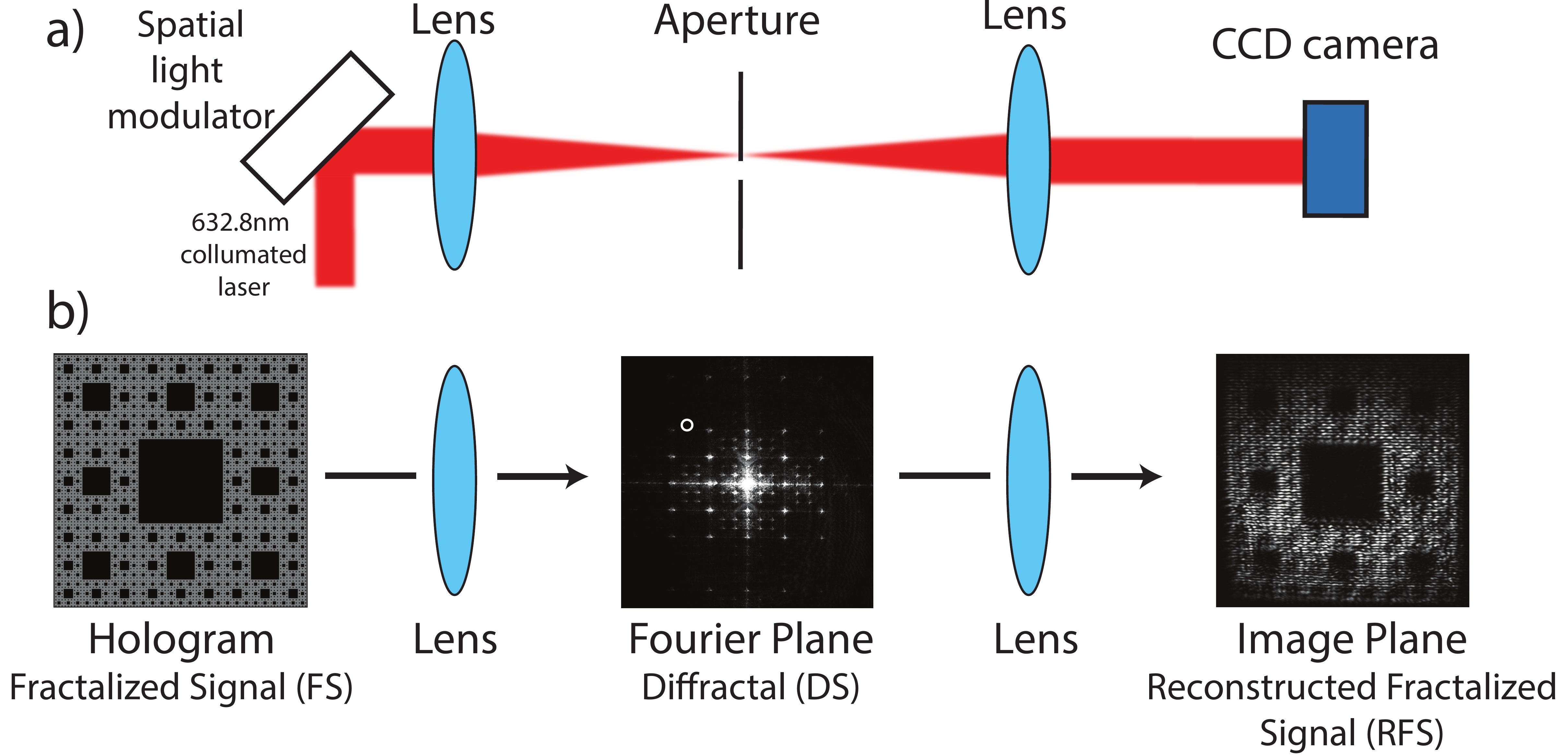}
\caption{a) Light from a $\lambda = $ 632.8-$nm$ wavelength laser is spatially filtered, expanded, and collimated and illuminates the full area of a 800x600 pixel spatial light modulator (SLM). The 4-$f$ system is composed of 2 5-$cm$ lenses placed after the SLM. The first lens Fourier transforms the fractal signal (FS) at the focal plane. An aperture is placed off-center at the focal plane and only transmits a portion of the diffractal (BDS). The second lens reconstructs the SLM image with the light that is transmitted through the aperture. b) The Sierpinski carpet hologram of order $n$= 5, CCD image in the focal plane of the hologram after the lens, and the reconstructed image. A circle denotes the area utilized to reconstruct the image.}
\label{ExpSetup}
\end{figure}

\indent We experimentally demonstrate the features of diffractals with a 4-$f$ optical arrangement where the 2-dimensional Fourier transform of a collimated fractal image, DS, lies in the focal plane of an imaging lens \cite{Goodman}.  The experimental setup that produces, filters and reconstructs FS is shown in Fig. \ref{ExpSetup}a) and experimentally reconstructed images, RFS, are shown in Fig. \ref{ExpSetup}b). An aperture of area approximately $0.8mm^2$ blocks the majority of DS. The placement of the aperture is shifted roughly $4mm$ horizontally and $5mm$ vertically from the central point, in the focal plane of the lens. The SLM image has a resolution of 800$\times$600 pixels (16mm$\times$12mm).  

When we employ diffractal architectures of high orders ($n>$4), we observe the phenomenon that is shown numerically: the placement of the aperture in DS is irrelevant in order to reconstruct the original image. When the aperture is moved laterally in the focal plane, RFS maintains strong resemblance to FS. In fact, only the intensity of RFS diminishes as the aperture translates further from the DS center; the outline remains fixed as the aperture moves.  For smaller fractal orders, RFS resembles FS only when the aperture is placed within $0.8mm$ of the center, where the spatial frequency components are concentrated. The signal-to-noise of the experiment and CCD camera sensitivity limit effective reconstruction, while the highest order $n$ is of FS is limited by the number of SLM pixels.  

\section{Discussion of Applications}\label{disc}
Here we make the distinction that diffractals are specific fractal structures. Not all fractals enable robust signal communications and the recursive transformation employed to generate FS and DS in Eqs.~\ref{BJ1} and~\ref{BF} differ fundamentally. For example, both of the Fourier Transform pairs FS and DS are fractals and carry iterated, self-similar features, but if we reverse their roles in our transmission system, the reconstruction will instead depend severely on the size and location of BDS. In fact, if in the example of the Sierpinski carpet, the center subsection becomes BDS, then the ROS will not resemble the OS, regardless of fractal order.  The diffractal is unique from general fractals and self-similar scale-invariance alone is an insufficient precondition for our system of robust reconstruction and spatial multiplexing. 

It may seem contradictory that higher-order fractals lead to more robust signal transmission since the finer structure of a higher-order fractal is itself harder to reconstruct. There are two perspectives of diffractals that explain the robust reconstruction. Firstly, the self-similar structures of higher-order fractals have a greater spatial frequency range and finer detail, and subsequently smaller BDS carry sufficient information to reconstruct OS. Secondly, the higher-order diffractals carry higher spatial-frequency components where the $k_x$ and $k_y$ components are coupled, and subsequently the location of the subsection in DS is unimportant. Here we have shown that BDS of arbitrary size and location carry sufficient information to reconstruct OS but in practice, there exists clear limitations for the robust reconstruction even in the limit of infinite-order FS.  

\begin{figure}[t!]
\includegraphics[width=\textwidth]{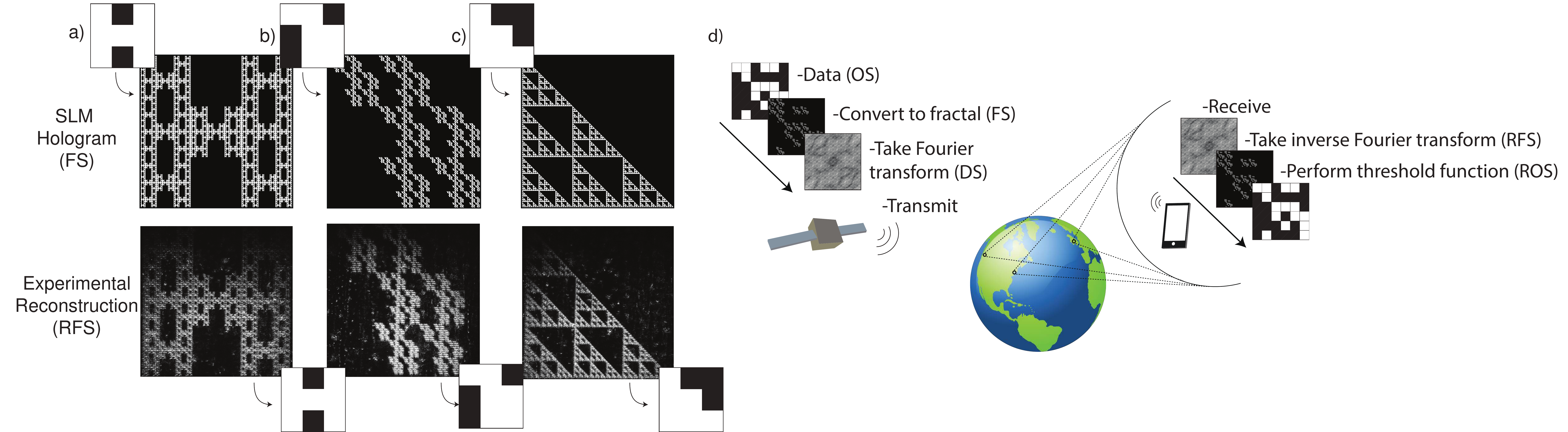}
\caption[]{Three examples of 512 9-bit spatial patterns, associated with base matrices 
a) $\Big[\begin{smallmatrix} 1 & 0 & 1\\ 1 & 1 & 1\\ 1 & 0 & 1 \end{smallmatrix}\Big]$, b) $\Big[\begin{smallmatrix} 1 & 1 & 0 \\ 0 & 1 & 1\\ 0 & 1 & 1\end{smallmatrix}\Big]$, and c) $\Big[\begin{smallmatrix} 1 & 0 & 0 \\ 1 & 1 & 0\\ 1 & 1 & 1\end{smallmatrix}\Big]$. The fractal signals FS are shown with their corresponding experimentally-reconstructed fractal signals RFS from the experimental setup in Fig. \ref{ExpSetup}a).  The lower-right inset shows the reconstructed original signal ROS. d) Example application: transmitted fractal signal FS is received at a far-field distance as a diffractal signal DS, where a roaming set of receivers, with only a diffractal subsection BDS, reconstructs the original signal OS.}
\label{Multiplex}
\end{figure}

There exists a trade-off with the diffractal architecture between robust reconstruction and high bit rate; a greater bit-rate is achieved with a larger base matrix, which can limit the maximum fractal order that is transmitted.  For example, with the $3\times 3$ or 9-element OS, there are 512 possible spatial bits, three of which are illustrated in Fig. \ref{Multiplex}a-c).  A $4\times 4$ base matrix requires $(\frac{4}{3})^{2n}$ more pixels than a $3\times 3$ to achieve the same fractal order, $n$. With a limited SLM pixel resolution, there is a choice between the generation of higher-order fractals and the utility of higher spatial bits.   

If the trade-off between bit-rate and robust reconstruction are mitigated, then the diffractal architecture could support a large number of roaming receivers with only one transmitter, such as wireless or satellite networks shown in Fig. \ref{Multiplex}d).  The self-similar properties of the diffractal architecture and their corresponding far-field pattern provide a method to reach a large number of receivers, possibly moving, without signal degradation.  The processing times required in the calculation of FS from OS are not trivial and scale with $r^{2n}$. Secondly, the refresh rates of a spatial light modulator or similar adaptive-optics device present constraints on the maximum achieved bit rate, which requires further consideration.  

\section{Conclusion}
We have shown that the diffractal architecture provides the beneficial features of spatial multiplexing and robust reconstruction. We have centered our demonstrations with the Sierpinski carpet, a familiar fractal, although our results could have been demonstrated with 511 other patterns similarly imprinted with the diffractal architecture. Data that is transmitted with the diffractal architecture is highly robust to intermediate-obstacle signal blocks and diffractal subsections of arbitrary size and location carry sufficient information to regenerate the original signal without distorting outlines of its pattern.  Our research illuminates potential applications in data transmission systems when one transmitter sends data to a large number of moving receivers or through noisy media.

The authors graciously acknowledge funding from NSF DMR 115-1783.
\end{document}